\documentclass[twocolumn,twoside,letterpaper,11pt]{article}
\usepackage{graphicx,xrrc,natbib}
\usepackage{times}
\begin{document}

\title{RADIO JETS IN COOLING CORES}   

\author{Jean A. Eilek}

\institute{New Mexico Tech}

\address{Socorro, NM 87801 USA}

\email{jeilek@aoc.nrao.edu}

\maketitle

\abstract

Almost every strong cooling core contains an active radio galaxy.
Combined radio and X-ray images reveal the dramatic interaction which
is taking place between the radio jet and the central cluster plasma.
At least two important questions can in principle be answered by
comparing the new data to theoretical models.  The first is how the
radio jet propagates, and disrupts, in the cooling core environment:
why are these cluster-center radio sources unusual?  The second is the
effect the radio jet has on the cooling core:  is it energetically
important to the core?  Thanks to the new data we are beginning to be
able to answer these questions. 	

\section{Introduction}

One topic of this meeting is the interaction of jets with their
environment.  This is an ideal area for the combination of radio
and X-ray observations.  I will focus on one particular aspect: 
the interaction of cooling cores in galaxy clusters with radio jets
that come from the massive black holes in the cluster center.  

\subsection{What is a cooling core?}

 We have known since the earliest
days of X-ray astronomy  that the atmosphere in some clusters
of galaxies has an unusually dense, X-ray bright core.  The high gas
density means a short cooling time.  If the thermal history of the
gas is simple -- if it is heated by the cosmological collapse of the
cluster, but subject to nothing else since that collapse -- the short
cooling time means hydrostatic equilibrium cannot be maintained in the
core. A slow, cooling-driven collapse will occur instead.  This has
been called a ``cooling flow''. The simplest interpretation
of the observations suggested that hundreds of solar masses per year are
collapsing into the cores of these clusters, with large amounts of 
cool to very cool gas residing in the inner tens of kpc.   

However, we now know things are not so simple.
With the advent of the newest X-ray satellites,
we now know that the large amounts of cool gas which
the simple models predicted are not there.  The dense cores of these
clusters are indeed a bit cooler than the rest of the atmosphere ({\it
e.g.} De Grandi \& Molendi 2002).  However, 
there is not nearly as much cold gas as
the early models predicted (Peterson {\it et~al.} 2004)
and certainly no evidence for smooth inflow.  In fact, CHANDRA images
of these clusters show that the inner tens of kpc have a
complex structure, not at all the spherically symmetric,
near-hydrostatic atmosphere that the simple cooling flow models
envisioned.

Thus, with little evidence for cooling and none for simple inflow,
  these dense cores need another name.  I call them 
``cooling cores''.\footnote{The name ``cool 
core'' might be even better; but 
in order to retain the connection to the older ``cooling flow'',
which is still in common use, I choose ``cooling core''}

Cooling cores are infrequent among
galaxy clusters.  They occur in unusually concentrated, massive clusters 
which seem to form in regions of high galaxy density (Loken, Melott \& Miller,
1999; Bullock {\it et~al.} 2001).  The high gas density in the cores
seems to be 
a consequence of this high concentration, which compresses large
amounts of intracluster gas into  quite small cores; the size of the
gas core is typically $\sim 100$-$150$ kpc.  The short
cooling time follows directly from this, and thus is 
due in the end to the happenstance of the cluster's formation.

\subsection{What is a cooling-core radio source?}

A massive, bright galaxy sits at the center of every cooling-core 
cluster.  Nearly all of these central galaxies contain an active 
galactic nucleus (AGN)
which currently supports a cluster-center radio source (CCRS) (initially
pointed out by Burns, 1990;  revisited by 
Markovi\'c 2004, also Eilek 2004a,b, ``E04''). 
While it has been suggested that this is due to the cooling
flow, it seems more likely due to the massive galaxy itself.  Ledlow
\& Owen (1996) show that the probability of a galaxy having a
radio source increases with galaxy size, so that galaxies as big as these
are very likely to have a radio source.

These CCRS are particularly interesting.  When we look at large samples of
CCRS, it becomes clear that they  are not typical
of the general radio source population.  
It must be that the unusual conditions in the cooling core affect the
CCRS in dramatic ways.  The CCRS in turn has a strong effect on the
gas in the cooling core, disturbing its simple equilibrium and possibly
providing significant heating to the gas. 

In this paper I explore both sides of this interaction.  I present an
overview of the data, and some theoretical work which may address
some of the important physics.  The keen reader can find more detail in
Eilek {\it et~al.} 2003, E04  and papers cited therein.

\section{The data reveal the interactions}

The place to start is with the data.  Samples with many objects
provide insight that observations of one or two sources cannot.  I have
formed two complete samples of nearby cooling-core clusters, one flux
limited, one volume limited (details in E04).  These
contain a total of 
41 cooling-core clusters with good information on the CCRS's. My sense of
the nature of CCRS's has come from these samples.
Another sample is that of Eilek \& Markovi\'c (2004; ``EM04''),
 which contains equal numbers (12 each) of rich clusters with and 
without cooling  cores. My insight into the nature of cooling cores 
comes mostly from this sample.
Finally, I have benefited from easy access to the data in two larger
samples:  the Owen-Ledlow sample of 250 radio galaxies in Abell clusters
(e.g., Owen \& Ledlow 1997), 
and the Ledlow {\it et~al.} (2004)
work on X-ray properties of 288 Abell clusters.

\subsection{The radio source disturbs the cooling core}

We know the CCRS disturbs the cooling core.   
This is apparent in essentially every CHANDRA image of a cooling core.  At
this writing, 10 of the 12 cooling cores in the EM04 sample have 
public-access CHANDRA data. {\it All} of these show disturbances
in the inner 20-40 kpc of the cooling core, which is also the scale of the
CCRS.    Some of the disturbed cooling cores
 are quite dramatic, with apparently spherical 
``bubbles'' and ``ghost cavities''.  Others are more subtle, with
asymmetric inner cooling cores,
 in which the RS appears to have displaced the ICM,
 or with complex interactions and apparent mixing 
between the RS and ICM.  In addition, faint, ring-like
features have been seen on larger scales in the X-ray gas of two 
clusters (Perseus, Fabian {\it et~al.} 2003;  and Virgo,   Forman {\it
et~al.} 2004).   These features  appear to be either sound waves or
shocks, centered on and driven by the AGN.

\subsection{The cooling core disturbs the radio source}

We also know the cooling core  disturbs the CCRS.  Nearly all (36 of 41)
cooling cores in the E04 sample contain detected radio sources;  I suspect
more will be detected with deeper observations.  Of these, 16 are large
and bright enough to be well imaged.  Very few of these 
are ``normal'' radio galaxies.  To explain this I need briefly to 
explain what a normal radio galaxy is. 

Almost all radio galaxies are jet-driven.  Their morphology is determined
by the interaction of the jet with its surroundings. In nearly every
case the directed jet continues from the galactic core all the way
to the extremities of the radio source, tens or hundreds of kpc out.  The
jet flow may continue directly to an outer hot spot,
as in classical double radio galaxies, or it may undergo a sudden transition
but continue on as a broad tail, as in tailed radio galaxies ({\it e.g.}, Eilek
{\it et~al.} 2003). 

\begin{figure}[t]
\begin{center}
\includegraphics[width=0.9\columnwidth]{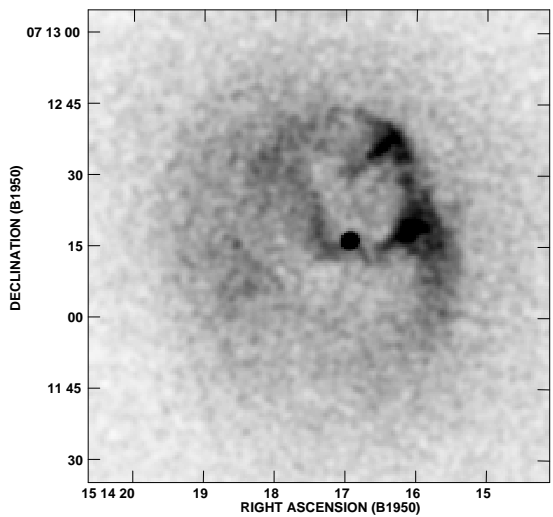}
    \caption{\small X-ray image of the cooling core A2052, linear extent
$\sim 60$ kpc.  The bright core, which hosts the AGN, is apparent in the
center of the figure, as are the ``bubbles'' which seem to have been evacuated
by the radio source.  Archival CHANDRA image, smoothed to 3.0 asec. 
See also Blanton {\it et~al.} (2003) for detailed analysis.}
    \label{fig:A2052X}
\end{center}
\end{figure}

\begin{figure}[t]
\begin{center}
\includegraphics[width=0.8\columnwidth]{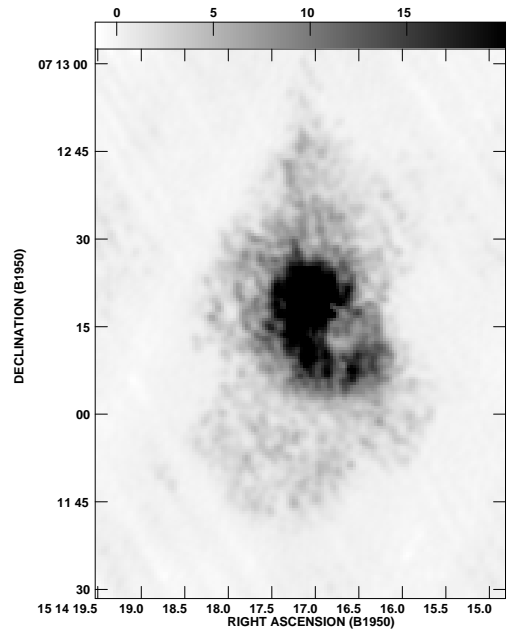}
    \caption{\small The radio source 3C317 in the core of A2052.  The AGN, 
which shows a pc-scale radio jet, is in the center of this image but burnt-out
in this exposure.  The scale of the image is similar to Fig.~\ref{fig:A2052X}.
The radio source appears to fill the cavities
apparent in Fig.~\ref{fig:A2052X} but also extends further to the
    north and south.
VLA image from Owen \& Ledlow (1997);  see also Zhou {\it et~al.} (1997) for
more detailed study.}
    \label{fig:3c317}
\end{center}
\end{figure}

By comparison, only two of the E04 CCRS sample are ``normal'' radio
galaxies.  These are Cyg A and Hyd A, two of the most poweful sources
in the nearby radio sky.  The rest of the well-imaged CCRS in this set
are diffuse,
amorphous sources.  They have a radio-loud core, which 
tells us the AGN is currently active,   but collimated jets exist only on 
kpc or  sub-kpc scales. It seems that the jet is disrupted close
to the galactic core, but its energy flow continues in a 
less collimated manner into the ambient cluster gas, creating a
diffuse, amorphous halo. 

\subsection{Three examples}

I illustrate with three examples, each showing a different type of
interaction between the radio source and the cooling core.  I show
the radio and X-ray images separately, to show both structures clearly.
Each source illustrated has a similar spatial scale, extending $\sim$
30-40 kpc from the core. 

\begin{figure}[t]
\begin{center}
\includegraphics[width=0.9\columnwidth]{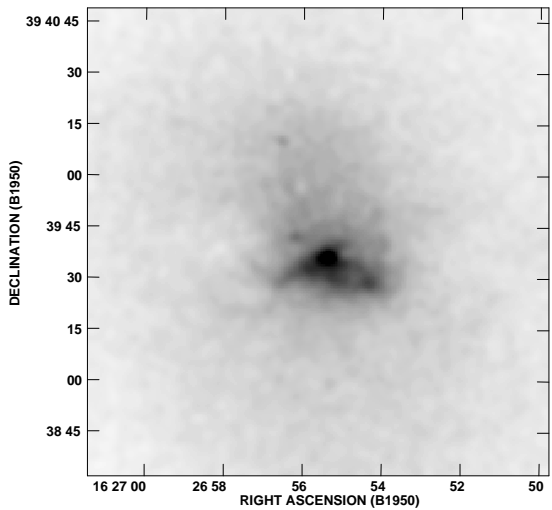}
    \caption{\small X-ray image of the cooling core in A2199, with linear
extent $\sim 75$ kpc north-south. The AGN coincides with the X-ray bright
spot in the center of the image.  The inner X-ray core is quite asymmetric
about this core, with an elongation to the north, a bright layer just to
the south, and faint ``cavitites'' to the east and west of the core. Archival
CHANDRA image, smoothed to 1.5 asec. See also Owen \& Eilek (1998), Johnstone
{\it et~al.} (2004).}
    \label{fig:A2199X}
\end{center}
\end{figure}

\begin{figure}[t]
\begin{center}
\includegraphics[width=\columnwidth]{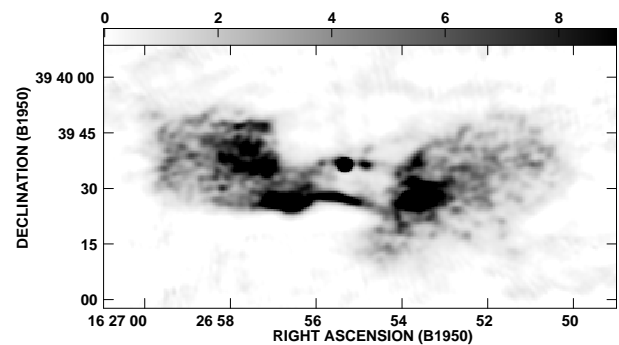}
    \caption{\small The radio source 3C338 in the core of A2199. This
north-south extent of this image is about half that of Fig.~\ref{fig:A2199X};
no radio emission has been detected on larger scales.  The
AGN, which coincides with the bright spot at the center of the image, contains
a two-sided, pc-scale jet which connects to the small kpc-scale jets visible
to the east and west of the core.  The radio-bright filament to the south,
which appears jet-like, does not coincide with any feature in the galaxy.
VLA image from Owen \& Ledlow (1997);  see also Owen \& Eilek (1998).}
    \label{fig:3C338}
\end{center}
\end{figure}

{\bf 3C317 in A2052} is a striking example of cavities in the X-ray loud gas
(apparent in Fig.~1)  which have been created by the radio source (shown
in Fig.~2). The clear cavities tell us that the X-ray and radio plasmas
are separated rather than well mixed.
 The brighter north-south regions of the radio halo coincide
with the X-ray cavities, but the radio halo also extends beyond the
cavity edges, into the less disturbed regions of the cooling core.  This
radio source has a strong radio core and a pc-scale jet.  No
jet has been detected on VLA scales (kpc or greater), but the bright radio
``tail'' to the south is suggestive of a semi-collimated flow from the
core into the radio halo.

{\bf 3C338 in A2199} is an unsual source in both radio and X-rays.  Figs.~3
and 4 show that the radio source and inner X-ray plasma  
seem to avoid each other.  The X-ray
gas shows an enhanced region to the north of the AGN, while the radio source
is offset to the south of the core.  The radio source again has active,
two-sided pc-scale jets, which continue to kpc scales and then grow very faint.
The region between the core and the bright radio filament to the south
is filled (at least in projection)
with enhanced X-ray emission.

{\bf M87 in the Virgo cluster} is an example of a more complex interaction
between the radio and X-ray plasmas.  Figure 5 shows that the X-rays display
a bright core, centered on the galactic nucleus, and also two bright
extended ``tails'' which lie to the west and southeast.  The radio
image shows a very bright inner core (which contains the famous jet, to which
I return in \S 4), and bright ``tails'' to the west and southeast, all within
a larger, amorphous halo. Both the radio and X-ray ``tails'' are suggestive
of flows within the halo;  however they do not coincide in detail.
 We also know from radio work this halo has a well-defined edge, which
implies that the halo has expanded (rather than diffused) into the X-ray
core.  The lack of X-ray holes, however, tells us that there has been good
mixing of the radio and X-ray plasmas across the edges of the radio halo.

\subsection{Is this the whole story?}

Can we generalize from these examples to all CCRS?  Two points come to
mind. 

First, it has been suggested that these core-halo
sources are the projection of a  normal, jetted radio galaxy
seen end-on.  However, the frequency of such sources in the CCRS
population, and the fact that the 4 (out of more than 150
well-imaged) Owen-Ledlow
sources which are amorphous {\it are in cooling cores}, argues against
this being just a projection effect.  
It is also worth noting that many of the CCRS which appear to have normal 
tails in published images, turn out in deeper images to have a diffuse halo
 (examples are A2597, Clarke 2004, or A2029, Markovi\'c 2004).

Second, the rest of the CCRS sample (20/36) are small, faint objects.  
Some are unresolved by Owen \& Ledlow (1997); some are resolved but
are too small and faint to be well imaged; and images of some  are only
available in the low-resolution NRAO VLA Sky Survey (Condon {\it et~al.}
1988).  These could be young sources, which have not yet grown large and
bright, or they could be due to jets which are intrinsically
weak.  We don't know which is the case;  but looking ahead to my
speculations in \S 4, it's 
tempting  to identify these as young, restarted jets, and suggest that
their larger-scale structure from previous activity cycles has faded.

\section{How does the radio jet affect the cooling core?}

Disturbances of the cooling core by the CCRS are common.
We know why the disruption occurs:  mass and energy
from the radio jet are slamming into the gas of the cooling core.
What we don't 
understand is the energetics of the system.  What is the jet power?
How important is this to to the energetics of the gas in the cooling core?
Because we can't measure the jet power directly, we must turn to models.
In this section I describe two types of models, which together can
constrain the jet power. 

\begin{figure}[t]
\begin{center}
\includegraphics*[width=0.8\columnwidth]{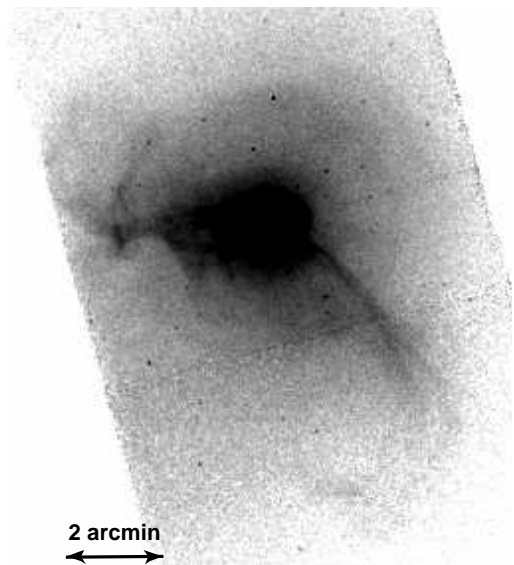}
    \caption{\small X-ray image of the core of the Virgo cluster.  The
north-south extent of the image $\sim 80$ kpc.  The AGN again sits in the
central X-ray peak;  striking features include the ``tails'' to the 
east and the southwest, and the ring-like structures centered on the core.
Complex smaller-scale structure exists in the inner core, which is burnt
out in this image.  See Forman {\it et~al.}  (2004), also Kraft 
{\it et~al.} (2004), for more details. CHANDRA image from Forman {\it et~al.} (2004).}
    \label{fig:VirgoX}
\end{center}
\end{figure}

\begin{figure}[t]
\begin{center}
\includegraphics[width=\columnwidth]{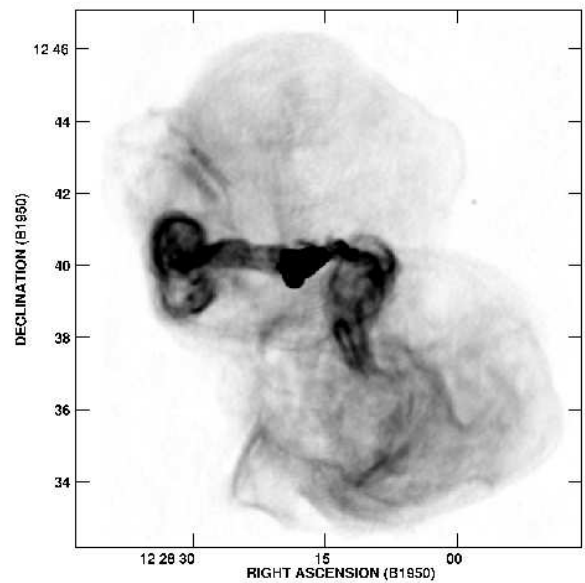}
    \caption{\small The M87 radio halo, in the core of the Virgo cluster.
The north-south extent of the image $\sim 80$ kpc, similar to Fig.~\ref{fig:VirgoX}.
The AGN and well-known kpc-scale radio jet are within the burnt out region in
the center of the image (but shown in Fig.~7).
 The ``tails'' to the east and the south, which appear
to be flows in the radio-loud plasma, coincide approximately with the X-ray
tails seen in Fig.~\ref{fig:VirgoX}. VLA image from Owen, Eilek \& Kassim (2000).}
    \label{fig:M87}
\end{center}
\end{figure}

\subsection{The general population:  dynamics?}

One approach uses simple dynamical models to relate the 
size of the source to the jet power and the source age.  This can be applied
to sources with good enough radio and X-ray data to understand the morphology
of the radio source and the structure of the local X-ray gas.

This method should be thought of as a toy model, because we do not
really know the three-dimensional structure of a given source
(although  we know CCRS as a  class are not just ``normal'' radio
sources seen in projection).     The radio haloes and 
X-ray cavities must be inflated against the pressure of the ambient gas.
The simplest model involves ongoing energy input to a quasi-spherical
``bubble'' which expands due to its own internal energy ({\it e.g.},
Fabian {\it et~al.} 2002, E04).
At late times  buoyancy will also contribute to, or even dominate,
 the growth of the source (Churazov {\it et~al.} 2001).  
Similar models can be developed for tailed sources; but since
only a few CCRS are clearly tailed these models have not yet been worked out
in as much detail (E04)

The key result is that these models are degenerate in 
jet power and source age.  The size of a driven bubble  measures the 
total energy input over the life of the source,
 $\int P_j dt$.  If we want to learn the jet power from this method
we need a separate estimate of the source age. There are at least two
approaches here, depending on one's taste.

One approach, traditional in radio astronomy, uses 
the observed radio spectrum and simple synchrotron physics to estimate the
age of the source.  This leads to quite short estimates for the source
age, $1$-$10$ Myr for our 3 example sources.  However, 
any {\it in situ} acceleration of the electrons  
can offset radiative losses and make the source appear ``younger'' than it 
is.  Thus, this approach only tells gives us a  lower limit to
the true source age, and a ``maximal'' jet power.

A more attractive approach  comes from simple dynamics ({\it e.g.}, Fabian 
{\it et~al.}  2002), and applies to sources for which we have temperature information
for the edges of the X-ray cavities.  If these shells are cooler
than the local ICM, we know they are not shocks, and  the expansion
speed of the bubble or cavity must be subsonic.  This translates to a lower
limit on the source age, and an upper limit on the jet power.  When
applied to our three example sources, this method suggests
ages no shorter than $ 10$-$30$ Myr, and jet powers no larger than $
4$-$70 \times 10^{44}$erg/s.  For comparison, the X-ray powers of the
cooling cores in these three objects are no larger than
 $\sim 0.3$-$1 \times 10^{44}$erg/s (using
the maximal cooling cores from Peres {\it et~al.}~1998).  Thus, it seems very likely
that the jet power is significant to the energetics of these cooling cores 
-- a conclusion which isn't surprising given the strong disturbance of the
cooling-core gas apparent in Figures \ref{fig:A2052X}, \ref{fig:A2199X}
and \ref{fig:VirgoX}.

\subsection{The general population:  radio power}

Dynamical models require good radio and X-ray images of a given cluster
core, which are not always available.  At least half of the CCRS sample
are too small or too faint to have good radio images.  We can only measure
the total radio power, $P_{\nu}$, for these sources.  We know from
synchrotron theory that $P_{\nu}$ is a highly imperfect tracer of the
underlying plasma energetics; so we must turn to statistics. If
a CCRS evolves according to the simple dynamical models described above,
we can use standard synchrotron analysis to predict how its radio
power evolves with time. These models show that 
the radio power depends only on the fraction of the jet power carried in
electrons.  We can thus  predict the mean ratio of radio power to jet 
electron power, over a sample of CCRS. 
Applying this to the E04 set of cooling cores, I find that this ratio
is small. The core X-ray power exceeds the jet's electron power by a
large factor;  
the electrons are not energetically important by themselves.

What this method lacks is any way to estimate the total jet power.
We must remember that the jet power can be carried by plasma and also
by magnetic field,  and that the plasma can be a mix of relativistic
leptons, relativistic ions and cooler, thermal gas.  It
is interesting to remember  that in our galaxy the  energy in
the relativistic ion component of cosmic rays exceeds that in electrons
by a factor $\sim 100$.   If this is also true for radio jets, then
many jets in the CCRS sample will have total power comparable to the
X-ray power of the cooling core in which they sit.

\section{How does the cooling core affect the radio jet?}

We know that radio jets in cooling cores have a good chance of
being disturbed close to their origin.
What we don't know is why this happens.  Indeed,
 we hardly know why jets in ``normal'' radio galaxies
stay stable and propagate as they do.   It 
may be that CCRS can be a useful example for understanding the larger questions
of jet stability and AGN duty cycles.  In this section I'll indulge
in some speculations on the physics of these unusual sources. 

\begin{figure}[t]
\begin{center}
\includegraphics[width=0.95\columnwidth]{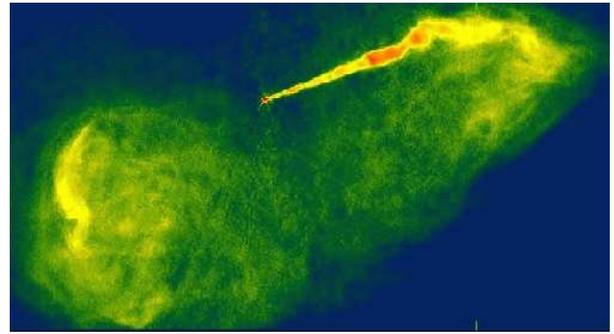}
    \caption{\small The inner radio core of M87, which appears as the
    burnt-out region in Fig.~6. This
VLA image from Hines, Owen \& Eilek (1989) shows the jet beginning to
twist and disrupt at a projected distance $\sim 3$ kpc from the
    core. Estimates of the jet angle to the line of sight range from
    $\sim 20^{\circ}$ to $\sim 40^{\circ}$, based on proper motions
    within the jet and kinematics of the nuclear gas disk ({\it e.g.},
    Biretta {\it et~al.} 1999, Ford {\it et~al.} 1994).}
    \label{fig:M87inner}
\end{center}
\end{figure}

\subsection{Hints from the data}

Over half of the well-imaged CCRS are not jet-driven on large scales.  
They do show a radio-loud core and pc-scale jets, which tells us
the AGN in these sources is currently active.  However, unlike most
radio galaxies, 
the well-collimated jet does not continue past kpc scales in these 
objects (this is illustrated by the 3 examples in \S2.3).  
This  suggests that  features which appear to be tails are, in
reality, more like the broad flows within the M87 halo (as in 
Fig.~\ref{fig:M87}).  It follows that that, although the jet disrupts
dramatically within a kpc or so of the AGN, the energy and mass flows continue
into the CCRS.  The radio jet and inner halo of M87 (shown in Fig.~7)
may be an example of this.  The jet is initially very well collimated,
but after a few kpc it  begins to bend and disrupt. 

The environment in which a CCRS jet propagates is also unusual.  Most
radio galaxies arise from normal bright ellipticals.  In these galaxies the
gas core extends only $\sim 1$ kpc ({\it e.g.}, Brighenti \& Mathews 1997),
 beyond which the jet propagates into
a low density, low pressure region.  In cooling cores, however, the
jet must fight its way out through a much larger core.  The 
dense, high-pressure region in cooling cores extends
 out to $\sim 50$-$100$ kpc
({\it e.g.} EM04).  These unusual conditions may make the jet more
susceptible to disruptive instabilities.

\subsection{Hints from simple models} 

CCRS statistics tell us that nearly all cooling cores -- and
probably all if we look hard enough -- contain an AGN which is currently
``on''.  But the simple models, described in \S 3.1, suggest
that these sources are unlikely to be older than $\sim 100$ Myr.  
These two statements can be reconciled only if the jet power fluctuates
on a similar timescale.  We might imagine, for instance, that the
jet maintains a steady power for the time needed to develop a large
radio halo, such as M87 or 3C317;  then goes into a low-power stage,
during which the radio halo fades or disperses;  then goes into another
high-power stage.   Two things follow from this.

(1) The small, faint cluster-core radio
sources are probably young and recently restarted.  
Their haloes from previous cycles must have faded. While such
restarting has been suggested occasionally for an individual radio source,
this is the first strong evidence of which I'm aware that such restarting
may be common.

(2) The large-scale  cluster-core
radio haloes must disappear quickly.  This is a long-standing
problem for the general radio galaxy population, where long synchrotron
lifetimes for the extended emission should keep a radio galaxy visible long
after its jet has turned off.  More rapid fading  might be possible in
these dense cluster cores, for two reasons.  The higher core pressure
makes higher magnetic fields likely, thus causing more rapid
synchrotron aging for the relativistic electrons.  In addition, the 
small turbulent scales likely in these cores could result in more rapid
turbulent dissipation of the magnetic field.   Both effects would
conspire to kill the extended radio emission fairly quickly once the
driving jet turns off.

\subsection{Hints from theoretical modelling}

There are some results in the literature which give us clues on  the
jet disruption and its ability to heat the cooling core. 

\begin{figure}[t]
\begin{center}
\includegraphics[width=0.9\columnwidth]{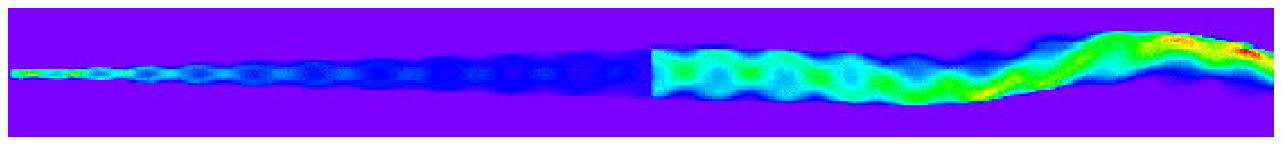}
\includegraphics[width=0.9\columnwidth]{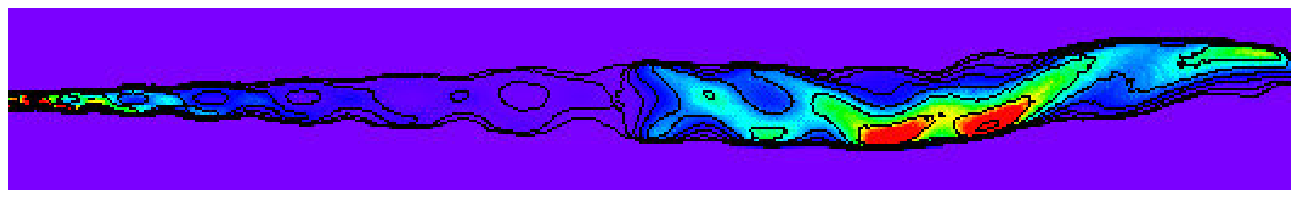}
\includegraphics[width=0.9\columnwidth]{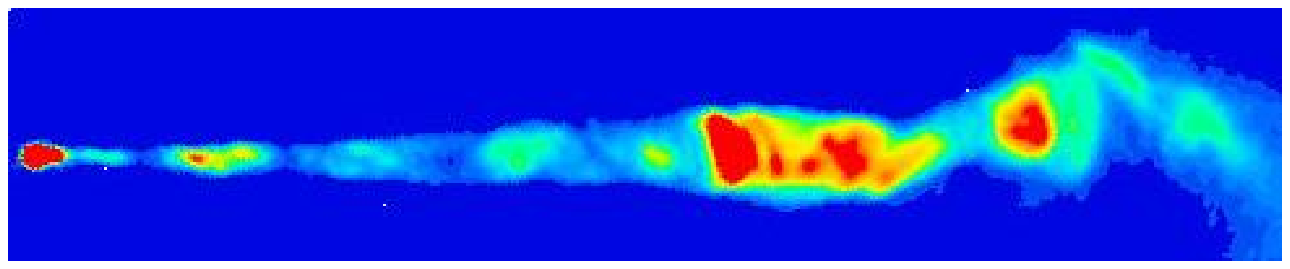}
 \caption{\small Comparing semi-analytic instability analysis of the
M87 jet to the real thing.  Top: a pseudo-synchrotron model of the
jet, seen in the plane of the sky.  The model assumes that helical and
elliptical surface Kelvin-Helmoltz modes are excited, and that the pressure and
wavelength of the modes jump midway down the jet.
Center, the same model jet, seen at a viewing angle of
$40^{\circ}$. Both from Hardee, Eilek \& Lobanov, work in progress.
Bottom, 15 GHz VLA image of the jet, from Owen, Hardee \& Cornwell (1989).}
    \label{fig:M87sim}
\end{center}
\end{figure}

First, analyses of a jet's Kelvin Helmholtz stability such as those 
of Hardee ({\it e.g.}, 2000, 2003, and references therein) provide
insight.  This work suggests that jet disruption happens
when the instability develops to the point of allowing significant
mass entrainment from the surroundings.  Hardee also notes that jets
which generate a 
lower density lobe or sheath surrounding the jet flow should be
expected to remain fairly stable. It may be that the extended
high-density region of a cooling core prevents the formation of
this stabilizing outer layer (Hardee, private communication).

Second, numerical simulations suggest that jets propagating into
steep density  ramps tend to destabilize.  Rizza {\it et~al.} (2000)
demonstrated this with  simulations of a jet
propagating into a classical cooling flow atmosphere (see also Loken
{\it et~al.}~1993).  We might expect such disordered jet flow to be effective
at heating the ambient gas in the cooling core, and simulations are
beginning to address this question.   Omma {\it et~al.} (2004) follow the
development of a high-momentum, jetted radio source through a cooling
core, with particular attention to its effect on the dynamical and
thermal state of the cluster gas (see also Basson \& Alexander 2003).
 Hughes, Hardee \& Eilek (work in progress) are undertaking
simulations of a perturbed jet propagating into a pressure ramp
with an inner core, in order to study the effect of the jet on the
cluster core as well as its stability.

Finally, it may be that we've caught one nearby jet
in the act of disrupting.  Lobanov, Hardee \& Eilek (2003) used
high-quality radio and optical images of the M87 jet to identify 
double-helical features in the inner jet which  which are consistent
with those caused by a Kelvin-Helmoltz instabilty. The dramatic twists
and bends in the outer jet, apparent in Fig.~7, may be the development
of the nonlinear stage of this instability. Hardee, Eilek \&
Lobanov (work in progress; Fig.~8) are extending this analysis to determine
whether physical conditions in the jet can be  consistent
with instability development as well as  other dynamic and observational
constraints known for this jet.

\section{Concluding Comments}

X-ray and radio data are converging to make this an exciting time
to be studying both cooling cores and the radio galaxies which sit at
their centers.  The new data raise at least as many  questions as
they answer, so we still have challenges ahead.

\section*{Acknowledgements}

Discussions with Frazer Owen, Tomislav Markovi\'c, Phil Hardee and Philip
Hughes have
been very helpful in this work.  I am also grateful to Mike Ledlow and
Tracy Clarke for sharing their unpublished data with me. 
Some of this work was done during my
sabbatical visits to the University of Oxford, and the Instituto di
Radioastronoma in Bologna;  I thank both institutions and the people in
them for their support. Finally, I want to thank the organizers of this
meeting for their financial support as well as for organizing a very
pleasant meeting.


\end{document}